\documentclass[runningheads]{llncs}
\usepackage[T1]{fontenc}
\usepackage{graphicx}
\newif\ifdraft
\draftfalse  

\ifdraft
  \usepackage[textsize=small]{todonotes}
\else
  \usepackage[disable]{todonotes}
\fi

\newif\ifinternaldraft
\internaldrafttrue

\usepackage[nostamp]{draftwatermark}
\usepackage{hyperref}
\hypersetup{
  pdftitle={Benchmarking Automated Knowledge Graph Construction from Semi-Structured Data — Preprint},
  pdfsubject={preprint; not peer reviewed}
}

\usepackage{tabularx}
\usepackage{amssymb}
\usepackage{booktabs}

\usepackage{array}
\usepackage{xcolor}
\usepackage{amsmath}
\usepackage{caption}
\newcommand{\bkr}[1]{\todo[color=purple!50,author=BKR,inline,size=\tiny]{#1}}
\newcommand{\tam}[2]{\todo[color=red!50,author=TAM,inline,size=\tiny]{#1}}

\begin{document}
\title{Benchmarking Automated Knowledge Graph Construction from Semi-Structured Data}
\titlerunning{Benchmarking Automated KG Construction from Semi-Structured Data}
%
\author{Tarek Al Mustafa\inst{1,2}\orcidID{0000-0001-7793-4483} \and 
Birgitta König-Ries\inst{2}\orcidID{0000-0002-2382-9722}}
\authorrunning{Al Mustafa et al.}
%
\institute{German Centre for Integrative Biodiversity Research Halle-Jena-Leipzig – iDiv, Puschstr. 4, 04103 Leipzig, Germany
\email{tarek.al\_mustafa@idiv.de} \and
Friedrich Schiller University Jena, Fürstengraben 1, 07743 Jena, Germany}
\maketitle              
%

\begin{center}
  \small\bfseries
  PREPRINT --- WORK IN PROGRESS\\
This manuscript represents an ongoing research project and may be revised substantially in subsequent versions.
\end{center}

\begin{abstract}
 Knowledge Graphs (KGs) play an increasingly important role in numerous applications ranging from traditional knowledge representation to serving as memory for LLMs to support downstream tasks.
 However, their construction is labor-intensive; thus, in recent years, numerous approaches for automizing this process have been proposed.
 Compared to the popularity of construction approaches that focus on textual input data, methods for semi-structured inputs remain underrepresented and as a result, no comprehensive benchmark and evaluation suite exists to judge the quality of mapping predictions and generated KGs.
 This is problematic, as a KG's quality has direct influence on the downstream applications it supports and thus, strong evaluation mechanisms for their construction are urgently needed.
In this work, we thus focus on the evaluation of KG construction from semi-structured data and present a benchmark and evaluation pipeline for KG construction that combines the quality dimensions (1) syntactic validity, (2) semantic accuracy, (3) consistency, (4) conciseness, (5) completeness, and (6) pragmatic quality measured on a KG's ability to provide answers to competency questions.
This work contributes a realistic task definition, extends current state of the art evaluation frameworks, allows evaluation of systems that predict mappings and RDF data alike, and combines evaluation of both KG construction and downstream usage.
We provide a comprehensive metrics suite, provide ten expert-curated datasets from seven domains, and showcase evaluation using two reference systems.

\end{abstract}
\section{Introduction}
\label{sec:intro}

Knowledge graphs (KGs) have become a prominent solution to represent data in a FAIR \cite{wilkinson2016fair} and machine-readable, -interpretable, and -actionable \cite{weiland2022fdo} format and thus increasingly provide the data foundation for downstream systems \cite{9416312,schneider2022decade}.
However, the difficulties of KG construction (domain modeling and mapping authoring) and usage (SPARQL\footnote{\url{https://www.w3.org/TR/sparql12-protocol/}} query formulation) continue to present the main bottlenecks for their wide-spread adoption, especially in domains outside of computer science.
Traditional methods for simplified KG construction (mapping languages and visual mapping tooling) have not yet been able to solve this issue \cite{duchateau2026grape,garcia2020shexml,junior2025protocolkgconstructiontasks,junior2017using,pinkel2014best,warren2024path}, but
continued improvements of machine-learning methods have led to the development of approaches that tackle this open problem with the use of LLMs \cite{ashiddiiqi2026reducing,hofer2024towards,schmidt2025llm} that might finally produce adequate solutions for users \cite{kim2025konda,stutz2024user}.
However, as the majority of systems target KG construction from unstructured textual data, research on KG construction from semi-structured data (tabular, JSON, XML) has produced few resources for streamlined evaluation.
This presents an open problem since semi structured formats are important sources for KG construction \cite{hofer2024construction} and highly relevant for research data \cite{broman2018data,crystal2022enabling,hollmann2025accurate}.
Similarly to Heist et al. \cite{heist2023kgreat}, we argue that evaluation frameworks should additionally consider the usefulness of generated KGs for downstream applications; but unfortunately, existing resources do not yet incorporate this quality dimension.
We therefore propose an evaluation framework and benchmark dataset that builds on existing approaches to support evaluation, comparison, and further development of KG construction systems.
Next to providing a unified evaluation framework for a variety of approaches, this resource will enable further analyses of increasingly relevant research questions: How well can automated approaches solve the KG construction task? What are common error sources of automated pipelines? To which degree are human-in-the-loop systems still required?
We thus make the following contributions:
\begin{description}
    \item \textbf{C1:} A task definition for KG construction from semi-structured data with required and optional input components that reflects realistic scenarios. (Section \ref{sec:task})
    \item \textbf{C2:} A review of the most relevant existing evaluation frameworks and how our approach extends them to evaluate the KG creation task at three stages -- mappings, materialized KG, and downstream task usefulness. (Section \ref{sec:related})
    \item \textbf{C3:} 
    An evaluation framework that supports two evaluation scenarios in which the system's choice of ontologies is (1) restricted to target ontologies to compare to reference mappings and (2) unrestricted.
    The framework is capable of scoring systems that produce mappings or RDF data across six quality dimensions -- syntactic validity, semantic accuracy (only for scenario 1), consistency, conciseness, completeness, and pragmatic quality.
    
    \bkr{Funktioniert das auch, wenn die Ontologien sehr gross sind? CHEBI oder so?}
    
    \tam{ja, dauert aber länger. die gesamte ontologie checken wir wahrscheinlich nur ein mal beim erstellen der datensätze, danach bei der eval berechnen wir nur relevante teile der ontos die im KG schema benutzt werden. dadurch dass man das framework lokal ausfüht hat man nicht wirklich internetprobleme, kann dann nur dauern wenn ein schema über 30 ontologien geht, aber dann ist efficiency future work}
    
    \bkr{ist die Pipeline eine eigene Contribution oder gehört die nicht zum Framework?}
    
    \tam{hmm kann man framen wie man will, die pipeline ist das was die umsetzung der theoretischen überlegung erlaubt, und das ist nicht trivial weil die pipleline module auch forschung sind (zb, wie komme ich von mapping table zu RDF=)}
    
    \item \textbf{C4:} A modular evaluation framework building on BLINKG \cite{castedo2026blinkg}. (Section \ref{sec:framework})
    \tam{ab welchem punkt ist das hier keine extension zu blinkg mehr bzw. ab wann steht das nicht mehr im vordergrund? dadurch, dass wir ja auch das similarity scoring anpassen, benutzen wir ja auch nicht mehr die gleichen scripte wie BLINKG}
    
    \bkr{kann man es als compatibility with BLINK framen? }
    
    \item \textbf{C5:} 
    A benchmark dataset of 10 expert-curated, real world datasets from x domains -- ecology, biodiversity research, genomics, copolymer chemistry, ...,  and history.
    Each dataset contains base data, competency questions (CQs) and gold answers, target ontologies, and gold mappings. (Section \ref{sec:data})
    \item \textbf{C6:} A showcase evaluation using two recent reference systems along multiple evaluation scenarios and a discussion of their results and their implications. (Section \ref{sec:eval})
\end{description}

\section{Task Definition}
\label{sec:task}
Our task definition attempts a realistic representation of deployment conditions faced by users.
Current benchmarking scenarios tend to restrict the KG creation task, for example by leaving out research papers or ontology tutorials from the input data.
We argue that these scenarios do not accurately represent deployment conditions of the KG construction systems we aim to evaluate and improve, as in deployment, users are able to supply additional context information to increase the probability of task success.
Additionally, the inclusion of optional context files enables ablation studies to further improve KG construction systems.
We therefore declare the following required and optional sources as viable input:
\begin{description}
    \item \textbf{I1:} (Research) data and metadata in semi-structured format (e.g., CSVs, JSON).
    \item \textbf{I2:} A set of competency questions the resulting KG should be able to answer based on \textbf{I1}.
    We omit questions that require linkage of the KG to other, preexisting sources, if the link targets are not also available in \textbf{I1}. 
    \item \textbf{I3:} A gold set of answers for the competency questions in tabular format. 
    \item \textbf{I4:} A set of ontologies as mapping targets in .ttl\footnote{\url{https://www.w3.org/TR/turtle/}} format. 
    Depending on the evaluation scenario, these are either target ontologies supplied by benchmark datasets or supplied by candidate systems.
    \item \textbf{I5:} If available, an \textit{intent description} file for the input data, for example \textit{'This table contains plants and their measured phenology traits across years'}.
    \item \textbf{I6:} If available, documentation and usage tutorials for each target ontology.
    \item \textbf{I7:} If available, one or more research papers based on the input data.
\end{description}

With the aforementioned input sources, we evaluate the KG construction task at three stages:

\begin{description}
    \item [Stage 1 -- Mappings:] Can a system predict KG mappings that use valid references to the source data and ontologies?
    Is the resulting KG schema logically consistent and concise?
    For Scenario 1: can a system interpret input data such that it is able to construct a KG schema that accurately depicts entity types and their relationships from the source data using target ontologies?
    \item [Stage 2 -- Materialized KG:] Is the RDF data generated from the predicted mappings logically consistent, concise, and does it contain all input data? Are entities and relationships mapped to correct ontology terms?
    \item [Stage 3 -- KG Usefulness:] Can the generated KG be used to successfully execute downstream tasks?
    For the first benchmark release, we focus on a KG's ability to answer the CQs used for its creation as evidence for how well a KG represents its original source data and domain knowledge and thus might serve as the data foundation for downstream applications.
    Additionally, as CQs can be interpreted as the queries and information needs \cite{case2016looking} real users may have for the KG, this also presents evidence for a KG's usefulness for question answering.
    
\end{description}

To evaluate these stages, we reuse the intrinsic quality dimensions from Zaveri et al.'s literature review on quality assessment for linked data \cite{zaveri2016quality} and adapt pragmatic quality \cite{BURTONJONES200584} from the domain of ontology evaluation to score a KG depending on its ability to solve a downstream task:

\begin{description}
    \item[D1 -- Syntactic Validity:]
    \textit{ the degree to which an RDF document conforms to the specification of the serialization format} \cite{zaveri2016quality}.
    In this context, syntactic validity is especially concerned with mapping predictions, e.g whether referenced input sources, predicted ontology terms, constructed IRIs, or datatypes/literals are syntactically valid.
    \item [D2 -- Semantic Accuracy:] \textit{ the degree to which data values correctly represent real world facts} \cite{zaveri2016quality}.
    Note that for this task, we use the provided input data as proxies for \textit{real world facts}.
    \item [D3 -- Consistency:] \textit{ a knowledge base is free of (logical/formal) contradictions with respect to particular knowledge representation and inference mechanisms} \cite{zaveri2016quality}.
    \item [D4 -- Conciseness:] \textit{ minimization of redundancy of entities at the schema and the data level. Conciseness is classified into (i) intensional conciseness (schema level) which refers to the case when the data does not contain redundant schema elements (properties and classes) and (ii) extensional conciseness (data level) which refers to the case when the data does not contain redundant objects (instances)} \cite{zaveri2016quality}.
    \item [D5 -- Completeness:] \textit{degree to which all required information is present in a particular dataset} \cite{zaveri2016quality}. In this work, the referenced \textit{required information} is the original input data.
    \item [D6 -- Pragmatic Quality:] \textit{a KG's usefulness in solving a downstream task.} For the first benchmark release, this refers to a KG's ability to answer CQs.
\end{description}

\section{Related Work}
\label{sec:related}
In this section, we present existing evaluation resources for KG construction from semi-structured data and their evaluation dimensions. We discuss which of their aspects can be reused and extended and where they fall short for our purposes and extensions are necessary. 

\textbf{SemTab\footnote{\url{https://www.cs.ox.ac.uk/isg/challenges/sem-tab/}} \cite{jimenez2020semtab}}. The Semantic Web Challenge on Tabular Data to Knowledge Graph Matching has been a yearly challenge since 2019.
Systems aim to solve the core tasks of Cell-Entity-Annotation, Column-Type-Annotation, and Column-Property-Annotation.
Systems aim to optimize table understanding and matching to existing knowledge graphs.
The mapping layer is not evaluated in SemTab as systems do not produce explicit declarative mappings, however, we reuse aspects of its evaluation approach (e.g., hierarchical prediction scoring) for the evaluation scenario in which predicted KGs are compared to reference KGs.

\textbf{RODI \cite{pinkel2015rodi}} is a benchmark to evaluate relational database to ontology mapping generation along multiple scenarios.
It covers three domains: scientific conferences, oil and gas exploration, and geographical data and its core feature is a query-based evaluation: 
Given reference SQL queries and their query results, systems produce KG mappings for the scenario ontology and write SPARQL queries against the mapped data.
System scores are then computed by comparing SPARQL query answers to provided references.
As such, RODI mainly evaluates mapping systems using a task-based evaluation, as it assumes a KG that can answer representative queries represents its source well.

As discussed by the authors and in \cite{castedo2026blinkg}, RODI is limited by excluding complex data transformations and by not including XML and JSON data.
Further, RODI does not evaluate mappings directly, and it also does not evaluate the generated graph outside of its query-answering capabilities: 
A generated graph could contain unnecessary statements and still answer queries correctly.

\textbf{KGrEaT \cite{heist2023kgreat}} Heist et al. argue that while many papers claim that enhanced KGs lead to better downstream task performance, most of them do not actually evaluate this claim.
As such, they propose KGrEaT, a framework to evaluate KGs based on downstream task performance for classification, regression, clustering, document similarity, entity relatedness, semantic analogies, and recommendation.
The framework precomputes artifacts for downstream tasks on input KGs, maps KG entities to evaluation dataset entities (a step they recognize as a potential source of errors), and subsequently evaluates how input graphs perform at different tasks.
As stated in their future work section, KG question answering is yet to be integrated in the resource.

\textbf{BLINKG \cite{castedo2026blinkg}.} BLINKG is a recent benchmark (accepted at TGDK\footnote{\url{https://drops.dagstuhl.de/entities/journal/TGDK}} in March 2026) for LLM-integrated KG generation. 
It contributes three main features:
(1) A tabular representation for KG mappings that separates a system's semantic mapping capabilities from its ability to formulate syntactically correct mappings in a mapping language.
As different systems can output their mapping decisions in this tabular format, they become comparable and can be evaluated using a gold table.
This table contains columns to predict source data reference, ontology property, entity class, related entity class, IRI template, joins, datatypes, language tags, and transformation functions (for example for correct datetime formatting).
(2) BLINKG evaluates predicted tables not via direct matching, but by employing a set of similarity metrics (Levenshtein distance, cosine similarity on string embeddings, cosine similarity on ontology-driven verbalizations) counting those cell predictions as correct that exceed a predefined threshold, recognizing that semantically equivalent mappings remain difficult to evaluate beyond syntactic matching.
For this evaluation framework, we introduce additional metrics to mitigate false positives that BLINKG's similarity scoring risks in certain edge-cases.
(3) A feature taxonomy that breaks down the mapping task into 12 different features, evaluated across three scenarios with increasing complexity ranging from schema-close to schema-distant.
The third scenario is especially relevant as it is concerned with constructing a KG for public procurement, a real-world use case, in which mapping difficulty is expressed by input to schema distance, meaning that LLMs must recognize and instantiate ontology schema patterns to solve the mapping task.
We aim to extend this realistic use case by additionally providing systems with ontology documentation, as BLINKG's current evaluation does not allow us to judge accurately, whether an LLM produced incorrect mappings, or whether it did not \textit{'know'} that the target ontology provides a pattern for this exact case.

\begin{figure}[ht!]
  \centering
  \includegraphics[width=\linewidth]{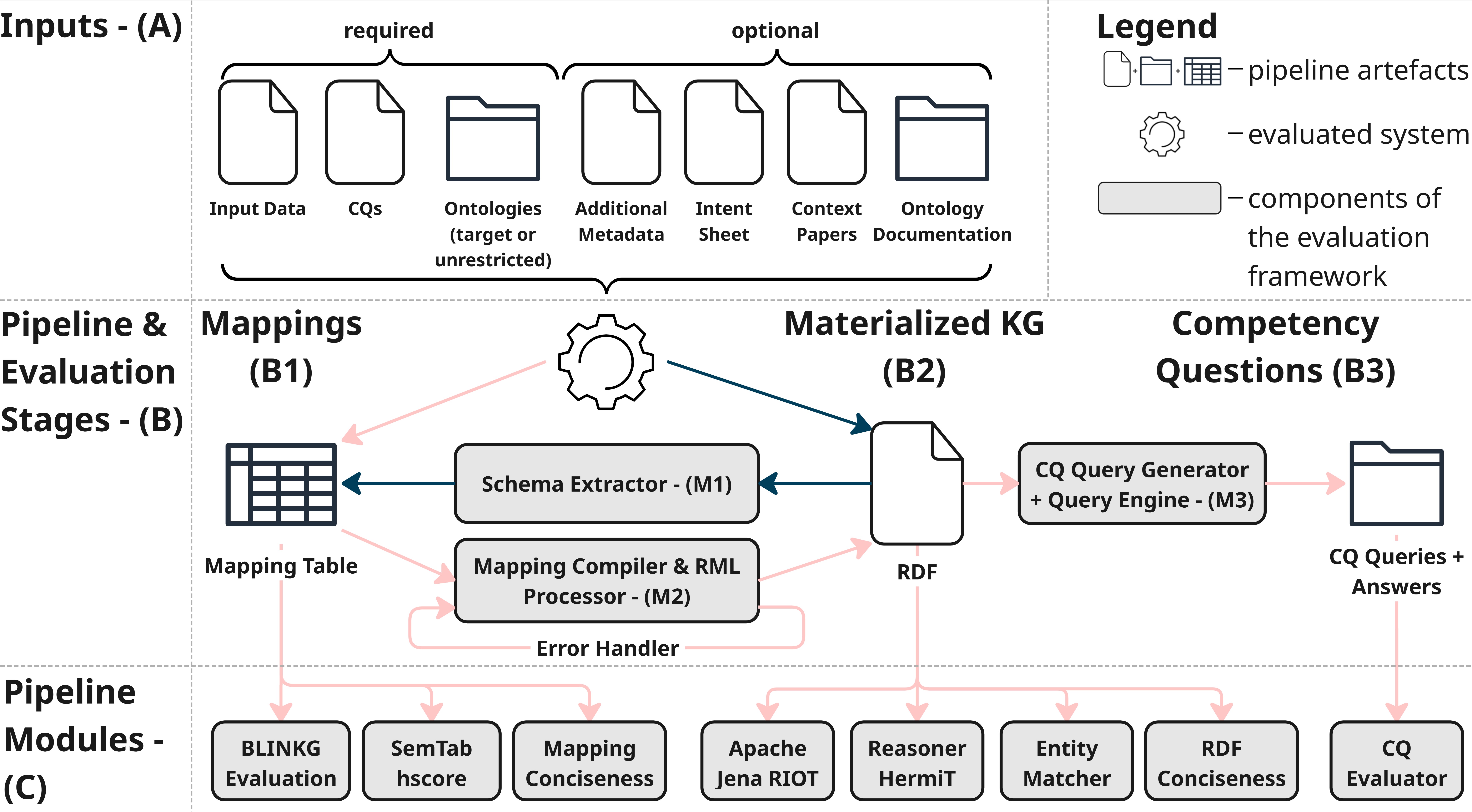}
  \caption{\textbf{Evaluation Pipeline Overview:}
  Section \textbf{(A)} lists all required and optional files that the pipeline accepts as valid input sources.
  These sources are made available to the evaluated system (cogwheel symbol), which either outputs a mapping prediction table (pink arrows) or a materialized KG as RDF data (blue arrows).
  Section \textbf{(B)} presents the three pipeline and evaluation stages -- Mappings \textbf{(B1)}, Materialized KG \textbf{(B2)} and Competency Questions \textbf{(B3)}  -- and section \textbf{(C)} lists the evaluation modules per stage.
  The pipeline facilitates KG construction and evaluation across stages by providing three processing modules.
  Following the blue arrows, the schema extraction module \textbf{(M1)} projects predicted RDF data into the format of a BLINKG mapping table to enable execution of the evaluation modules for the mapping stage.
  For systems that predict mapping tables directly (pink arrows), the second processing module \textbf{(M2)} compiles mapping predictions into RML mappings and processes them into RDF data.
  If processing fails, an error handler attempts to fix the mapping syntax and rerun processing.
  In cases where the error handler is not successful, the pipeline only reports metrics for \textbf{B1}.
  Finally, \textbf{(M3)} first takes RDF data as input and generates SPARQL queries for each provided competency question, and then processes these queries using a SPARQL engine to allow evaluation of the queries' results.
  }
  \label{fig:1}
\end{figure}

\section{Evaluation Framework} \label{sec:framework}

In this section, we present the design of our evaluation pipeline and evaluation metrics per quality dimension across stages.
\subsection{Evaluation Pipeline}
Fig. \ref{fig:1} depicts the design of the proposed evaluation pipeline.
Evaluated systems may produce either mapping predictions as a BLINKG-style mapping table, or output RDF data directly.
The pipeline provides processing modules \textbf{(M1-M3)} to connect these two entry points and derive the representations required by the evaluation stages (Mappings \textbf{(B1)}, Materialized KG  \textbf{(B2)}, and CQs \textbf{(B3)}).
Given RDF output, the schema extractor \textbf{(M1)} reconstructs a
BLINKG-compatible mapping table, allowing graph-producing systems to also be evaluated at the mapping level.
Conversely, the mapping compiler \textbf{(M2)} translates predicted mapping tables into executable RML mappings and materializes the resulting RDF graph.
Finally, the CQ processing module \textbf{(M3)} generates and executes SPARQL queries for the supplied competency questions.
The stage-specific evaluation modules shown in Section \textbf{(C)} score the metrics described in the following section.

\subsection{Quality Dimensions per Stage and Evaluation Metrics.}

In this section, we present the proposed metrics suite in detail.
For each quality dimension, we initially restate and expand upon the definition given in Sec. \ref{sec:task}, list the proposed metrics and how they should be interpreted, and then define each metric in detail.
We additionally refer readers to Table \ref{tab:metrics-overview} for a comprehensive overview.

\newcommand{\fullcalc}{\(\checkmark\)}
\newcommand{\partialcalc}{\((\checkmark)\)}
\newcommand{\nocalc}{\(\times\)}

\begin{table}[ht!]
\centering
\caption{Overview of evaluation metrics by quality dimension, evaluation
stage, and evaluation scenario.}
\label{tab:metrics-overview}
\footnotesize
\setlength{\tabcolsep}{3pt}
\renewcommand{\arraystretch}{1.02}

\begin{tabularx}{\textwidth}{
    @{}
    >{\raggedright\arraybackslash}p{0.17\textwidth}
    >{\centering\arraybackslash}p{0.065\textwidth}
    >{\raggedright\arraybackslash}X
    >{\centering\arraybackslash}p{0.070\textwidth}
    >{\centering\arraybackslash}p{0.085\textwidth}
    >{\centering\arraybackslash}p{0.100\textwidth}
    @{}
}
\toprule
\textbf{Dimension}
& \textbf{ID}
& \textbf{Metric}
& \textbf{Stage}
& \multicolumn{2}{c}{\textbf{Ontology Scenario}} \\
\cmidrule(lr){5-6}
&&&&
\textbf{Target}
& \textbf{Arbitrary} \\
\midrule

Syntactic validity
& SV1
& Criterion-level syntactic validity
& B1
& $\checkmark$
& $\checkmark$ \\

& SV2
& RDF serialization validity
& B2
& $\checkmark$
& $\checkmark$ \\
\hline
\addlinespace[2pt]
Semantic accuracy
& SA1
& Mapping-level semantic accuracy
& B1
& $\checkmark$
& $\times$ \\

& SA2
& Assertion-level semantic accuracy
& B2
& $\checkmark$
& $\times$ \\

& SA3
& Entailment-aware semantic accuracy
& B2
& $\checkmark$
& $\times$ \\

& SA4
& Entity-centred neighbourhood accuracy
& B2
& $\checkmark$
& $\times$ \\
\hline
\addlinespace[2pt]
Consistency
& CS1
& Criterion-level mapping consistency
& B1
& $\checkmark$
& $\checkmark$ \\

& CS2
& Materialized-KG consistency
& B2
& $\checkmark$
& $\checkmark$ \\
\hline
\addlinespace[2pt]
Conciseness
& CN1
& Intensional conciseness
& B1/B2
& $\checkmark$
& $(\checkmark)$ \\

& CN2
& Extensional conciseness
& B1/B2
& $\checkmark$
& $\checkmark$ \\

& CN3
& Identifier-collision freedom
& B1/B2
& $\checkmark$
& $\checkmark$ \\

& CN4
& Duplicate fact-generation freedom
& B1
& $\checkmark$
& $\checkmark$ \\

& CN5
& Property-value conciseness
& B1/B2
& $\checkmark$
& $\checkmark$ \\
\hline
\addlinespace[2pt]
Completeness
& CM1
& Source-reference completeness
& B1
& $\checkmark$
& $\checkmark$ \\

& CM2
& Source-information completeness
& B2
& $\checkmark$
& $\checkmark$ \\

& CM3
& Population completeness
& B2
& $\checkmark$
& $\checkmark$ \\
\hline
\addlinespace[2pt]
Pragmatic quality
& --
& Per-CQ answer quality
& B3
& $\checkmark$
& $\checkmark$ \\

& PQ1--PQ3
& Macro-averaged answer quality
& B3
& $\checkmark$
& $\checkmark$ \\

& PQ4
& Micro-averaged answer quality
& B3
& $\checkmark$
& $\checkmark$ \\

& PQ5
& Exact CQ satisfaction
& B3
& $\checkmark$
& $\checkmark$ \\

& PQ6
& Hierarchy-aware CQ answer quality
& B3
& $\checkmark$
& $(\checkmark)$ \\

& --
& Stratified evaluation
& B3
& $\checkmark$
& $\checkmark$ \\

\bottomrule
\end{tabularx}

\vspace{2pt}
\parbox{\textwidth}{\scriptsize
\textit{Stages:}
B1 = mappings; B2 = materialized KG; B3 = KG usefulness through
competency-question answering.
\textit{Ontology scenarios:}
Target = construction using benchmark-provided target ontologies and reference
mappings; Arbitrary = construction using ontologies supplied by the candidate
system.
$\checkmark$ indicates full calculation,
$(\checkmark)$ conditional or partial calculation, and
$\times$ that the metric is not calculated.
For arbitrary ontologies, CN1 is limited to equivalences expressed in the
submitted ontologies, while PQ6 requires an alignment between candidate and
gold ontology-term answers.
CN4 additionally requires mapping-execution provenance.
}
\end{table}

\subsubsection{D1 Syntactic Validity.}
Syntactic validity is defined as \textit{the degree to which an RDF document conforms to the specification of its serialization format} \cite{zaveri2016quality}.
For the KG construction task, we extend this notion to the mapping layer and evaluate whether mapping predictions form valid references to the provided source data and target vocabularies and use well-formed mapping components.

We define two metrics: criterion-level syntactic validity (SV1) evaluates the validity of source and ontology references, joins, IRI templates, datatypes, language tags, and transformation functions, independently of their semantic or logical correctness.
RDF serialization validity (SV2) evaluates whether the materialized RDF passes strict serialization validation.
High syntactic validity therefore indicates that a system produces well-formed and resolvable mapping components and RDF data, but does not imply that the chosen mappings are semantically accurate or logically consistent.

To evaluate syntactic validity of predicted mappings, we define \(\mathcal{C}^{SV}\) as the set of criteria similar to BLINKG's evaluation to score whether predictions exist in the data sources and target vocabularies:
\begin{description}
    \item[\(c_{\mathrm{src}}^{SV}\): Source reference existence --]
    Does the predicted data reference exist in the input source?

    \item[\(c_{\mathrm{class}}^{SV}\): Subject class existence  --]
    Is the predicted entity class a known class in the target ontologies?

    \item[\(c_{\mathrm{prop}}^{SV}\): Property existence  --]
    Is the predicted property a known property in the target ontologies?

    \item[\(c_{\mathrm{join}}^{SV}\): Join validity  --]
    Do the referenced join columns or paths exist?

    \item[\(c_{\mathrm{iri}}^{SV}\): IRI template validity  --]
    Do all template variables refer to existing source fields, and does the instantiated template produce valid IRIs?
    
    \item[\(c_{\mathrm{dtype}}^{SV}\): Datatype validity --] Is the predicted datatype a valid and recognized RDF or XML Schema datatype?

    \item[\(c_{\mathrm{lang}}^{SV}\): Language-tag validity --] Is the predicted language tag well formed according to the applicable language-tag syntax?

    \item [\(c_{\mathrm{func}}^{SV}\): Transformation function validity -- ] Does the predicted data transformation function exist?


\end{description}

\paragraph{SV1: Criterion-Level Syntactic Validity}
For each criterion \(c\), syntactic validity is measured as the proportion of applicable predictions that pass the check:
\[
\mathrm{SV}_{1}(c)
=
\frac{
    \text{no. of predictions passing criterion } c
}{
    \text{no. of predictions to which criterion } c \text{ applies}
}.
\]
We report both an exact score \(\mathrm{SV}_{1}^{\mathrm{raw}}(c)\) and a similarity-based score \(\mathrm{SV}_{1}^{\mathrm{sim}}(c)\).
The raw score only counts exact matches between raw predictions and source data/target ontologies (score = 0 otherwise), while the similarity-based score accepts predictions above a predefined threshold, reusing BLINKG's evaluation approach.
We report the criterion-level scores \(\mathrm{SV}_{1,\mathrm{micro}}^{m}(c)\) and their macro-average across all criteria with at least one applicable prediction as \(\mathrm{SV}_{1,\mathrm{macro}}^{m}\), for
\(m\in\{\mathrm{raw},\mathrm{sim}\}\).\footnote{Note that in BLINKG, the terms micro and macro are used to compare and average the performance of predictions across multiple runs for the same system to gauge whether a system requires multiple attempts for stability.}


\paragraph{SV2: RDF Serialization Validity.}
At the materialized-KG layer, RDF data is validated using Apache Jena RIOT\footnote{\url{https://jena.apache.org/documentation/io/}} with strict parsing and IRI and literal checking enabled.
RIOT distinguishes warnings, errors, and fatal errors.
We record their absolute numbers as $N_{\mathrm{warn}}$, $N_{\mathrm{err}}$, and $N_{\mathrm{fatal}}$, respectively, and define the total number of reported validation issues as
\(
N_{\mathrm{issue}}
=
N_{\mathrm{warn}}+
N_{\mathrm{err}}+
N_{\mathrm{fatal}}.
\)
Strict RDF serialization validity is defined as
\[
\mathrm{SV}_{2} =
\begin{cases}
1, &
N_{\mathrm{warn}} =
N_{\mathrm{err}} =
N_{\mathrm{fatal}} = 0,\\
0, & \text{otherwise}.
\end{cases}
\]
We retain the emitted diagnostics including severity and line and column positions.
We note that there may be more errors than the reported counts suggest, as fatal parsing errors may terminate validation before processing is completed.
\bkr{hattest Du da nochmal geguckt, was Warnings sind? Sind das issues, die wirklich stören?}
\tam{diese liste müsste die warnings ganz gut beschreiben: IRI violations, including both “Bad IRI” and “Unwise IRI” cases; invalid language tags; an rdf:langString literal without a language tag; an rdf:dirLangString literal without language and direction; a lexical form that is invalid for its declared datatype; some unrecognized or questionable node and Unicode cases. For example, an invalid lexical value such as "300"^^xsd:byte is reported as a warning rather than as an error. The code explicitly comments that IRI violations are handled as warnings during checking. The tokenizer also emits warnings for some illegal or questionable characters, including: certain prohibited characters in an IRI; control characters in IRIs; Unicode replacement characters in prefixed names or blank-node labels; Unicode non-characters in strings.}


\subsubsection{D2 Semantic Accuracy.}
Semantic accuracy is defined as \textit{the degree to which data values correctly represent real-world facts} \cite{zaveri2016quality}.
For the KG construction task, we operationalize this notion as whether the semantics assigned to the source data agree with the expert-curated reference mappings and KG.
As this comparison requires a shared set of target ontologies, semantic accuracy is evaluated only in Scenario~1.

We define four metrics: mapping-level semantic accuracy (SA1) compares predicted mapping components against the reference mappings; assertion-level semantic accuracy (SA2) evaluates the correctness of type and property assertions in the materialized KG; entailment-aware semantic accuracy (SA3)
accounts for assertions that are logically entailed rather than explicitly identical; and entity-centered neighborhood accuracy (SA4) evaluates the local
semantic representation of individual entities.
Exact, similarity-, hierarchy-, and entailment-aware variants account for different degrees of agreement with the reference.
High semantic accuracy therefore indicates that a system assigns semantics to the source data that closely agree with the benchmark's reference interpretation, but does not by itself imply that the resulting KG is logically consistent, complete, concise, or useful for downstream tasks.

While \textbf{D1} evaluates whether mapping predictions are valid references to the input data and target vocabularies, semantic accuracy evaluates whether the predicted mapping table agrees with the provided gold mapping table.
We therefore reuse BLINKG's evaluation for the following mapping fields:
\[
\mathcal{C}_{\mathrm{map}}^{SA}
=
\{
c_{\mathrm{src}}^{SA},
c_{\mathrm{prop}}^{SA},
c_{\mathrm{class}}^{SA},
c_{\mathrm{relclass}}^{SA},
c_{\mathrm{iri}}^{SA},
c_{\mathrm{join}}^{SA},
c_{\mathrm{dtype}}^{SA},
c_{\mathrm{lang}}^{SA},
c_{\mathrm{func}}^{SA},
c_{\mathrm{funcout}}^{SA}
\}.
\]
For every criterion \(c \in \mathcal{C}_{\mathrm{map}}\), we report BLINKG's exact and similarity-based precision, recall, and \(F_1\), denoted by the matching modes \(m\in\{\mathrm{raw},\mathrm{sim}\}\).
To reduce false positives, a similarity-based match is accepted only, if the predicted value does not already resolve to an incorrect valid reference and the gold value is its highest-scoring candidate.
Accepted matches receive full rather than proportional credit.
In addition, inspired by SemTab's hierarchy-aware correctness score \cite{jimenez2020results}, we introduce a third and fourth matching mode \(m= hierraw\) and \(m= hiersim\) for
\(\mathcal{C}_{\mathrm{hier}}^{SA}
=
\{
c_{\mathrm{class}}^{SA},
c_{\mathrm{relclass}}^{SA},
c_{\mathrm{prop}}^{SA}
\}\).
The function assigns partial credit to broader or narrower terms related through an ancestor-descendant path by combining a reciprocal-rank-inspired score decay based on prediction distance and a relative scaling factor that punishes wrong predictions on gold terms that are close to the root more than on gold terms far away from the root.
For a predicted term \(a\) and gold term \(g\), let \(d(a,g)\) denote their shortest hierarchical distance and let \(h(g)\) denote the shortest distance from the hierarchy root to \(g\).
We define their relative hierarchical distance as \(\delta(a,g)=\frac{d(a,g)}{\max(1,h(g))}.\)
To distinguish broader from narrower predictions, let
\(\lambda_{\mathrm{anc}}\) and \(\lambda_{\mathrm{desc}}\) denote
direction-specific penalty parameters, where
\(\lambda_{\mathrm{anc}} < \lambda_{\mathrm{desc}}\). We define
\[
\operatorname{hscore}(a,g)
=
\begin{cases}
1,
&
\text{if } a=g,
\\[1mm]
\dfrac{1}{
1+\lambda_{\mathrm{anc}}\delta(a,g)
},
&
\text{if \(a\) is an ancestor of \(g\),}
\\[4mm]
\dfrac{1}{
1+\lambda_{\mathrm{desc}}\delta(a,g)
},
&
\text{if \(a\) is a descendant of \(g\),}
\\[4mm]
0,
&
\text{otherwise}.
\end{cases}
\]
Following SemTab, these partial-credit scores replace the number of exact matches when calculating hierarchy-aware precision, recall, and \(F_1\).
Class predictions are compared using the transitive closure of \texttt{rdfs:subClassOf}, while property predictions are compared using \texttt{rdfs:subPropertyOf}.
Invalid ontology terms,
missing predictions, and properties of an incompatible kind receive a score of zero.
For each matching mode \(m\in\{\mathrm{raw},\mathrm{sim},\mathrm{hierraw},\mathrm{hiersim}\}\), we thus report criterion-level scores \(\mathrm{SA}_{1,micro}^{{m}}(c)\) and their macro-average across criteria as \(\mathrm{SA}_{1,macro}^{{m}}\).


Additionally, we declare the following metrics for evaluating semantic accuracy on materialized KGs:
\paragraph{SA2: Assertion-Level Semantic Accuracy} We evaluate two criteria:
\(c_{\mathrm{insttype}}\), whether generated entities have the correct \texttt{rdf:type} assertions, and \(c_{\mathrm{instprop}}\), whether aligned subjects and objects are connected by the correct properties.
We report exact and hierarchy-aware precision, recall, and \(F_1\), for \(m\in\{\mathrm{raw},\mathrm{hier}\}\).
Exact matching requires identical classes or properties, whereas hierarchy-aware matching replaces binary matches with the previously defined \(\operatorname{hscore}(a,g)\).
Where several assertions are available for the same entity or entity pair, predicted and gold assertions are aligned through maximum-weight one-to-one matching.
We denote the resulting criterion-level scores by \(\mathrm{SA}_{2,\mathrm{micro}}^{m}(c)\) and their macro-average across both criteria by \(\mathrm{SA}_{2,\mathrm{macro}}^{m}\).

\paragraph{SA3: Entailment-Aware Semantic Accuracy} Following semantic precision and recall for ontology alignment evaluation~\cite{euzenat2007semantic,fleischhacker2010practical}, we additionally evaluate whether predicted assertions are entailed by the gold KG and vice versa. 
Let \(K_P\) and \(K_G\) denote the predicted and gold-standard KGs, respectively, and let \(O\) denote the set of target ontologies.
Additionally, let \(T_P\) and \(T_G\) denote the evaluated type and property assertions in \(K_P\) and \(K_G\), respectively:
\[
P_{\models}
=
\frac{
    \left|
    \left\{
        t\in T_P
        \mid
        K_G\cup O \models t
    \right\}
    \right|
}{
    |T_P|
},
\qquad
R_{\models}
=
\frac{
    \left|
    \left\{
        t\in T_G
        \mid
        K_P\cup O \models t
    \right\}
    \right|
}{
    |T_G|
}.
\]
We report \(\mathrm{SA}_{3}^{P}=P_{\models}\), \(\mathrm{SA}_{3}^{R}=R_{\models}\), and their harmonic mean \(\mathrm{SA}_{3}^{F_1}=F_{1,\models}\).
Entailments are computed under the same reasoning regime as D3.
To prevent trivial entailment from an inconsistent knowledge base, predicted KGs found to be inconsistent receive an entailment-aware score of zero.

\paragraph{SA4: Entity-Centered Neighborhood Accuracy} Global assertion-level scores may be dominated by entities with many assertions.
We therefore additionally compare the local neighborhoods of individual entities, following RDF comparison approaches based on the incoming and outgoing statements of mapped nodes \cite{almeida2015approxmap,tzitzikas2012blank}.
Let \(N_P(e)\) and \(N_G(e)\) denote the one-hop incoming and outgoing type and property assertions of entity \(e\) in the predicted and gold KGs.
Entity-centered neighborhood accuracy is the macro-average of the resulting local \(F_1\) scores:
\[
\mathrm{SA}_{4}^{m}
=
\frac{1}{|E|}
\sum_{e\in E}
F_1^{m}
\left(
    N_P(e),
    N_G(e)
\right),
\qquad
m\in\{\mathrm{raw},\mathrm{hier}\}.
\]
Here, \(E\) contains matched entities as well as unmatched predicted and gold entities, with unmatched entities receiving a score of zero. 
Consequently, every entity receives equal weight irrespective of its number of assertions.

\textbf{Scenario 2:} For the first pipeline release, we omit measuring semantic accuracy for the second evaluation scenario in which no target ontologies or reference mappings are provided, as we are not able to deterministically score whether predicted terms are semantically accurate representations of the base data.
As this is still an open research question, we aim to add evaluation once appropriate solutions exist.

\subsubsection{D3 Consistency.}
Consistency refers to whether a knowledge base is free of logical or formal contradictions with respect to the applied knowledge representation and inference mechanisms \cite{zaveri2016quality}.
For the KG construction task, we operationalize this notion as a gold-independent assessment of whether predicted mappings and their materialized RDF are compatible with the axioms of the target ontologies.

We define two metrics: criterion-level mapping consistency (CS1) evaluates whether mapping decisions are compatible with class disjointness, property domains and ranges, datatypes, property kinds, and language-tag constraints; materialized KG consistency (CS2) evaluates the resulting knowledge base using an OWL reasoner and records whether it is logically consistent.
Mapping-level checks distinguish compatible, incompatible, and unknown cases, since the absence of an entailment does not constitute a violation under the open-world assumption.
High consistency therefore indicates that the predicted representation does not contradict the supplied ontology axioms, but does not imply that the chosen classes and properties are the semantically most appropriate ones or that the resulting KG is complete, concise, or useful for downstream tasks.

\bkr{Das funktioniert ja nur dann, wenn die target ontologies konsistent waren. Sollten wir irgendwo erwähnen, dass wir das überprüfen (im dataset kapitel)}

At the mapping layer, we evaluate \(\mathcal{C}^{CS}\) as the set of criteria:
\begin{description}
    \item[\(c_{\mathrm{disjoint}}^{CS}\): Type compatibility --]
    Do explicit or inferred class assignments place an entity in disjoint classes?
    
    \bkr{wissen wir das hier schon? wir haben bisher doch nur mappings (also schema level, oder? das passiert doch in B1? wäre die Frage dann nicht eher: erzeugen wir klassen, die per definition keine individuen haben können?}

    \item[\(c_{\mathrm{domain}}^{CS}\): Domain compatibility --]
    Is the predicted entity class compatible with the domain of the
    predicted property?

    \item[\(c_{\mathrm{range}}^{CS}\): Object-range compatibility --]
    Is the predicted related entity class compatible with the range of
    the predicted object property?

    \item[\(c_{\mathrm{dtype}}^{CS}\): Datatype compatibility --]
    Is the predicted datatype compatible with the range of the predicted
    datatype property?

    \item[\(c_{\mathrm{pkind}}^{CS}\): Property-kind compatibility --]
    Does an object property map to an entity and a datatype property to
    a literal?

    \item[\(c_{\mathrm{lang}}^{CS}\): Language-tag compatibility --]
    Is a language tag used only for a literal with a compatible
    string-like representation?
\end{description}
Because the absence of an entailment does not constitute a violation under the open-world assumption, each applicable prediction is assigned a status from \(\mathcal{S} = \{\text{compatible},\text{incompatible},\text{unknown}\}\).
For domain and range checks, a prediction is compatible if the predicted class is equal to or a subclass of the declared domain or range, incompatible if the resulting type assignment conflicts with an ontology axiom such as class disjointness, and unknown if neither compatibility nor incompatibility can be established.

\paragraph{CS1: Criterion-Level Consistency}
For each criterion \(c \in \mathcal{C}^{CS}\) and status \(s \in \mathcal{S}\), we report the proportion of applicable predictions assigned that status:
\[
\mathrm{CS}_{1,\mathrm{micro}}^{s}(c)
=
\frac{
    n_s(c)
}{
    n_{\mathrm{applicable}}(c)
}.
\]
We report these criterion-level rates together with their macro-averages across all criteria with at least one applicable prediction, denoted by \(\mathrm{CS}_{1,\mathrm{macro}}^{s}\).
We also retain the absolute numbers of incompatible and unknown predictions.
To perform these checks, the pipeline computes the transitive subclass and subproperty closures, inherited property domains and ranges, and class-disjointness relations over the target ontologies.

For materialized KGs, the RDF graph is combined with the target ontologies and evaluated using the HermiT reasoner \cite{glimm2014hermit} under OWL~2 Direct Semantics \cite{horrocks2012owl}.
We separately record whether the reasoning process completes successfully, $\mathrm{RunSuccess}\in\{0,1\}$, and whether the resulting knowledge base is logically consistent as \(CS_2 = 1\) if yes, and \(CS_2 = 0\) if the reasoner runs successfully but the KG is not consistent.
If reasoning fails because of a timeout, unsupported input, or a reasoner exception, \(CS_2\) is left undefined and the corresponding execution status is reported. 
For consistent knowledge bases, we additionally report the number of unsatisfiable named classes, excluding \texttt{owl:Nothing}, as $N_{\mathrm{unsat}}$.
We also derive error counts from criterion-level checks:
For every consistency criterion \(c \in \mathcal{C}^{CS} \), we report the absolute numbers $N_{\mathrm{inc}}(c)$ and $N_{\mathrm{unknown}}(c)$ of incompatible and undetermined assertions, together with their normalized rates $\mathrm{CS}^{s}_{2,\mathrm{micro}}(c)$. 
We additionally report the total number of criterion violations, $N_{\mathrm{inc,total}}=\sum_{c \in \mathcal{C}^{CS}} N_{\mathrm{inc}}(c)$, and the number $N_{\mathrm{affected}}$ of unique assertions involved in at least one incompatibility.
The latter prevents assertions that violate several criteria from being interpreted as several independent errors.

\subsubsection{D4 Conciseness.}
Conciseness concerns the minimization of redundant representations in a KG.
Following \cite{zaveri2016quality}, we distinguish between intensional conciseness, which concerns redundant schema elements, and extensional conciseness, which concerns redundant representations of entities.
For the KG construction task, we extend this notion to redundancy introduced by mapping decisions, identifiers, and generated property values, and evaluate conciseness at both the mapping and materialized-KG stages.

We define five metrics: intensional conciseness (CN1) evaluates redundant equivalent classes and properties; extensional conciseness (CN2) detects multiple identifiers representing the same entity; identifier-collision freedom (CN3) detects the inverse case in which one identifier represents multiple distinct entities; duplicate fact-generation freedom (CN4) evaluates whether mapping constructs redundantly generate identical RDF statements; and property-value conciseness (CN5) detects multiple equivalent representations of the same property value.
High conciseness therefore indicates that a system represents the required information without unnecessary duplication or conflation, but does not imply that the represented information is semantically accurate, complete, logically consistent, or useful for downstream tasks.
Mapping-level conciseness is only fully observable for systems that provide mapping predictions, as some forms of generation redundancy cannot be reconstructed from materialized RDF alone.

Based on the metrics compiled from \cite{FurberHepp2011SWIQA,kontokostas2014test,lei2007framework,mendes2012sieve,ruckhaus2013analyzing}, we evaluate five forms of redundancy.
Let \(\ell \in \{\mathrm{B1},\mathrm{B2}\}\) denote the evaluation stage.
For \(\ell=\mathrm{B1}\), conciseness metrics measure which entities and identifiers the predicted mappings imply and for \(\ell=\mathrm{B2}\), conciseness metrics evaluate the actually materialized KG.
Note that mapping-level conciseness is only fully observable for systems that provide mapping predictions, as not all relevant information for the defined conciseness metrics can be inferred from a materialized KG.

\paragraph{CN1: Intensional Conciseness.} Let
\(T^{\mathrm{CN}}_{\mathrm{schema}}=\{\mathrm{class},\mathrm{prop}\}\) denote the evaluated schema-element types.
For each \(t \in T^{\mathrm{CN}}_{\mathrm{schema}}\), let \(S^\ell_t\) be the set of distinct class or property identifiers used at stage \(\ell\), and let \(\equiv_t\) denote the equivalence relation induced by the reflexive, symmetric, and transitive closure of the corresponding ontology axioms and benchmark-provided equivalence declarations.
We define
\[
\mathrm{CN}^{\ell}_{1}(t)
=
\frac{
    \left|S^\ell_t/{\equiv_t}\right|
}{
    \left|S^\ell_t\right|
}.
\]
Here, \(S^\ell_t/{\equiv_t}\) denotes the quotient set of schema identifiers under \(\equiv_t\). 
We report the scores separately for classes (\(\mathrm{CN}^{\ell}_{1,\mathrm{class}}\)) and properties (\(\mathrm{CN}^{\ell}_{1,\mathrm{prop}}\)), as well as their micro average with equal weights across all elements (\(\mathrm{CN}^{\ell}_{1,\mathrm{micro}}\)) and macro average where both types are weighted equally (\(\mathrm{CN}^{\ell}_{1,\mathrm{macro}}\)).
Schema elements are treated as redundant only when their equivalence is established by the target ontologies or explicitly declared by the benchmark.
Related but non-equivalent ontology terms are not collapsed.

\paragraph{CN2: Extensional Conciseness.} Let \(E_G\) denote the set of gold entities, let \(E^\ell_P\) denote the candidate entity representations at stage \(\ell\), and let
\(A^\ell \subseteq E^\ell_P \times E_G\) be the candidate-to-gold entity alignment.
For each gold entity \(e \in E_G\), we define its set of candidate representations as
\(
R^\ell(e)
=
\left\{
r \in E^\ell_P
\mid
(r,e) \in A^\ell
\right\}.
\)
Extensional conciseness is then defined as
\[
\mathrm{CN}^{\ell}_{2}
=
\frac{
    \left|
    \left\{
    e \in E_G
    \mid
    |R^\ell(e)|>0
    \right\}
    \right|
}{
    \sum_{e \in E_G}|R^\ell(e)|
}.
\]
Gold entities without a candidate representation are excluded from the numerator and denominator because their absence is evaluated under completeness.
We additionally report \(N^\ell_{\mathrm{\text{excessabs}}}\) as the absolute number of duplicate identifiers  and \(N^\ell_{\mathrm{\text{excessent}}}\) as the number of gold entities represented by more than one identifier.

\paragraph{CN3: Identifier-Collision Freedom.}
Whereas CN2 detects cases in which one gold entity is represented by multiple candidate identifiers, CN3 detects the inverse case in which one candidate identifier represents multiple distinct gold entities.
For each candidate representation \(r \in E^\ell_P\), let \(G^\ell(r)=\left\{e \in E_G\mid(r,e) \in A^\ell\right\}\) denote the set of gold entities represented by \(r\).
Identifier-collision freedom is defined as
\[
\mathrm{CN}^{\ell}_{3}
=
1-
\frac{
    \left|
    \left\{
    r \in E^\ell_P
    \mid
    |G^\ell(r)|>1
    \right\}
    \right|
}{
    \left|
    \left\{
    r \in E^\ell_P
    \mid
    |G^\ell(r)|>0
    \right\}
    \right|
}.
\]
Candidate identifiers that cannot be aligned to a gold entity are excluded because this metric specifically evaluates the conflation of known distinct entities. 
The metric detects, for example, collisions caused by IRI templates whose variables do not uniquely identify source entities.
We additionally report \(N^\ell_{\mathrm{collisionabs}}\) as the total number of excess gold entities merged into candidate identifiers and \(N^\ell_{\mathrm{collisionids}}\) as the number of candidate identifiers representing more than one gold entity.

\paragraph{CN4: Duplicate Fact-Generation Freedom.}
CN4 evaluates whether different executions of the predicted mapping constructs redundantly generate the same RDF statement.
Let \(\widetilde{T}^{\mathrm{B1}}_P\) denote the multiset of statement-generation events produced by the predicted mappings before RDF duplicate elimination, and let \(\operatorname{supp}(\widetilde{T}^{\mathrm{B1}}_P)\) denote the corresponding set of distinct RDF statements.
We define
\[
\mathrm{CN}^{\mathrm{B1}}_{4}
=
\frac{
    \left|
    \operatorname{supp}(\widetilde{T}^{\mathrm{B1}}_P)
    \right|
}{
    \left|
    \widetilde{T}^{\mathrm{B1}}_P
    \right|
}.
\]
For RDF datasets containing named graphs, statements are compared as quads, such that the graph identifier contributes to statement identity.
We additionally report \(N^{\mathrm{B1}}_{\mathrm{duplicatefacts}}\) as the absolute number of redundant statement-generation events.
CN4 is a mapping-level metric because it evaluates redundancy caused by the predicted mapping constructs and requires the generation provenance retained before RDF duplicate elimination.
It is not measured on the materialized KG, where identical RDF statements may no longer be distinguishable due to mapping processor behavior.

\paragraph{CN5: Property-Value Conciseness.}
Following and extending the property-level conciseness measure proposed by \cite{mendes2012sieve}, CN5 evaluates whether the same subject is assigned multiple equivalent representations of a property value. 
Let \(V^\ell(s,p)\) denote the values associated with subject \(s\) and property \(p\) at stage \(\ell\). For \(\ell=\mathrm{B1}\), \(V^{\mathrm{B1}}(s,p)\) is a multiset in which repeated mapping-generation events are retained; for \(\ell=\mathrm{B2}\), \(V^{\mathrm{B2}}(s,p)\) contains the distinct RDF values actually present in the materialized KG.
Let \(\equiv_V\) denote value equivalence after benchmark-defined literal canonicalization and entity-identity normalization, and let \(Q^\ell(s,p)=\left\{[v]_{\equiv_V}\mid v \in V^\ell(s,p)\right\}\)
denote the set of value-equivalence classes represented for a subject-property pair.
For each property \(p\), let \(S^\ell_p=\left\{s\mid|V^\ell(s,p)|>0\right\}\) denote the subjects with at least one value for \(p\).
Property-value conciseness is defined as
\[
\mathrm{CN}^{\ell}_{5}(p)
=
\frac{
    \left|
    \left\{
    s \in S^\ell_p
    \;\middle|\;
    |V^\ell(s,p)|
    =
    |Q^\ell(s,p)|
    \right\}
    \right|
}{
    |S^\ell_p|
}.
\]
A subject is therefore concise with respect to \(p\) if each generated or
materialized value belongs to a different equivalence class. We report the
property-level scores \(\mathrm{CN}^{\ell}_{5}(p)\), their micro average with
equal weights across all applicable subject-property pairs
\(\mathrm{CN}^{\ell}_{5,\mathrm{micro}}\), and their macro average with equal
weights across all applicable properties
\(\mathrm{CN}^{\ell}_{5,\mathrm{macro}}\). We additionally report
\(N^\ell_{\mathrm{valueabs}}\) as the total number of excess equivalent value
representations and \(N^\ell_{\mathrm{valuepairs}}\) as the number of
subject-property pairs containing at least one such redundancy.

\subsubsection{D5 Completeness.}
Completeness refers to the degree to which all information required for a given task is represented \cite{zaveri2016quality}.
For the KG construction task, we operationalize this notion with respect to the benchmark's source data and assume a closed-world setting in which the provided source data delimit the information that may be required.
We consider source information required if it is either needed to derive the gold-standard answer to at least one competency question or represented by the benchmark's reference mappings or reference KG.
Completeness is therefore evaluated against these ontology-independent source references and information units rather than against the particular ontology terms or graph patterns used in the reference representation.
\begin{sloppypar}
We reuse and adapt three completeness metrics compiled from \cite{feeney2014improving,FurberHepp2011SWIQA,mendes2012sieve,paulheim2014improving}:
source-reference completeness (CM1) evaluates whether all required source fields are used by the predicted mappings; source-information completeness (CM2) evaluates whether the values and relations contained in the source data are represented in the materialized KG; and population completeness (CM3) evaluates whether all source entities are represented in the KG.
High completeness therefore indicates that a system preserves the information contained in the source data, but does not imply that this information is represented using semantically accurate ontology terms, without redundancy, or in a form that is useful for downstream tasks.
\end{sloppypar}

\paragraph{CM1: Source-Reference Completeness.}
Source-reference completeness measures whether all required source fields are used by the predicted mappings.
Let \(F_G\) denote the benchmark-defined set of required source references and let \(F_P^{\mathrm{B1}}\) denote the source references used by at least one predicted mapping statement.
We define
\[
\mathrm{CM}^{\mathrm{B1}}_{1}
=
\frac{
    \left|F_P^{\mathrm{B1}}\cap F_G\right|
}{
    |F_G|
}.
\]
A source reference is considered represented if it participates in at least one applicable subject map, object map, join, identifier template, or transformation.
We additionally report \(N^{\mathrm{B1}}_{\mathrm{referencemiss}}\) as the absolute number of required source references not used by the predicted mappings.
Where a benchmark contains several source files or structures, we report source-level scores as well as their micro average over all required references and macro average across source files.

\paragraph{CM2: Source-Information Completeness.}
Source-information completeness measures whether the values and relations contained in the source data are represented in the materialized KG, independently of the ontology terms or graph patterns used to represent them.
Let \(U_G\) denote the benchmark-defined set of required source information units. A literal-valued information unit has the form \(u=(e,f,v),\) where \(e\in E_G\) is a source entity, \(f\in F_G\) is a source reference, and \(v\) is its normalized source value.
An entity-valued information unit has the form \(u=(e_s,f,e_o),\) where \(e_s,e_o\in E_G\) denote the source subject and object entities.

Let \(\operatorname{repr}^{\mathrm{B2}}_P(u)\in\{0,1\}\) indicate whether information unit \(u\) is represented in the predicted KG.
For a literal-valued unit \((e,f,v)\), the indicator equals one if a value equivalent to \(v\) occurs in a candidate subgraph anchored at a candidate representation of \(e\).
For an entity-valued unit \((e_s,f,e_o)\), it equals
one if candidate representations of \(e_s\) and \(e_o\) are connected within the corresponding candidate subgraph.
The connecting representation may consist of a direct property assertion or an arbitrary-length path containing intermediate entities.
For each required source reference \(f\), let \(U_G(f)=\left\{u\in U_G\mid u\text{ originates from }f\right\}.\)
We define source-information completeness for \(f\) as
\[
\mathrm{CM}^{\mathrm{B2}}_{2}(f)
=
\frac{
    \displaystyle
    \sum_{u\in U_G(f)}
    \operatorname{repr}^{\mathrm{B2}}_P(u)
}{
    |U_G(f)|
}.
\]
We report the source-reference-level scores \(\mathrm{CM}^{\mathrm{B2}}_{2}(f)\), their micro average with equal weights across all required source information units \(\bigl(\mathrm{CM}^{\mathrm{B2}}_{2,\mathrm{micro}}\bigr)\), and their macro average with equal weights across all applicable source references \(\bigl(\mathrm{CM}^{\mathrm{B2}}_{2,\mathrm{macro}}\bigr)\). 
We additionally report \(N^{\mathrm{B2}}_{\mathrm{informationmiss}}\) as the absolute number of missing source information units.

\paragraph{CM3: Population Completeness.}
Population completeness measures the proportion of source entities represented in the materialized KG.
Reusing the source entity set \(E_G\) and the representation function \(R^\ell(e)\) defined for CN2, we define
\[
\mathrm{CM}^{\mathrm{B2}}_{3}
=
\frac{
    \left|
    \left\{
    e\in E_G
    \mid
    |R^{\mathrm{B2}}(e)|>0
    \right\}
    \right|
}{
    |E_G|
}.
\]
Unlike CN2, the denominator contains all source entities, such that entities without a candidate representation reduce the score.
We additionally report \( N^{\mathrm{B2}}_{\mathrm{entitymiss}}\) as the absolute number of missing source entities.
CM1 is evaluated at the mapping stage and CM2/CM3 are evaluated on the materialized KG.

\subsubsection{D6 Pragmatic Quality.}
Pragmatic quality concerns the usefulness of a KG for solving a downstream task.
For the first benchmark release, we operationalize this notion through CQ answering:
a KG is considered pragmatically useful to the extent that it enables the information needs expressed by the benchmark's CQs to be answered correctly and completely.
We therefore evaluate the answers obtained from the generated KG against the provided gold-standard CQ answers rather than requiring agreement with a particular graph representation.

We report per-CQ answer correctness, completeness, and F1 and aggregate them using macro-averaged (PQ1--PQ3) and micro-averaged answer quality (PQ4).
Exact CQ satisfaction (PQ5) measures the proportion of questions answered completely correctly, while hierarchy-aware answer quality (PQ6) gives partial credit to semantically related ontology-term answers where applicable.
We additionally report results stratified by CQ characteristics such as answer cardinality, query complexity, mapping challenge, and reasoning requirements.
High pragmatic quality therefore indicates that the generated KG supports the information needs represented by the benchmark's CQs, but does not imply that the KG is intrinsically correct or useful for downstream tasks not covered by those questions.

Let \(Q\) denote the benchmark set of CQs.
For each CQ \(q \in Q\), let \(A_{q'}\) denote the answer set returned by executing the pipeline generated query \(q'\) that corresponds to \(q\) on the evaluated KG, and let \(G_q\) denote the corresponding gold-standard answer set.
Before comparison, answer tuples are canonicalized to account for equivalent representations of IRIs and literals, and duplicate answer tuples are removed.
Boolean and scalar answers are represented as singleton answer sets.
If query execution fails, the answer-quality scores for \(q\) are set to zero.

\paragraph{Per-CQ Answer Quality}
For each competency question \(q \in Q\), answer correctness and completeness are defined as
\[
\operatorname{Correctness}(q)
=
\frac{
    \left|A_{q'} \cap G_q\right|
}{
    \left|A_{q'}\right|
},
\qquad
\operatorname{Completeness}(q)
=
\frac{
    \left|A_{q'} \cap G_q\right|
}{
    \left|G_q\right|
}.
\]
A failed query receives a score of zero.
If both the returned and gold-standard answer sets are empty, both scores are defined as one; if only one of them is empty, both scores are defined as zero.

For each competency question, \(F_1(q)\) is calculated as the harmonic mean of answer correctness and completeness.

\paragraph{PQ1--PQ3: Macro-Averaged Answer Quality}
We report macro-averaged correctness, completeness, and \(F_1\) over all
competency questions, with failed queries contributing a score of zero:
\[
\mathrm{PQ}_{1}
=
\operatorname{Correctness}_{\mathrm{macro}},
\qquad
\mathrm{PQ}_{2}
=
\operatorname{Completeness}_{\mathrm{macro}},
\qquad
\mathrm{PQ}_{3}
=
F_{1,\mathrm{macro}}.
\]
\paragraph{PQ4: Micro-Averaged Answer Quality}
We additionally report micro-averaged correctness, completeness, and \(F_1\) over all answer tuples:

\[
\begin{gathered}
\mathrm{PQ}_{4}^{\mathrm{Correctness}}
=
\operatorname{Correctness}_{\mathrm{micro}},
\qquad
\mathrm{PQ}_{4}^{\mathrm{Completeness}}
=
\operatorname{Completeness}_{\mathrm{micro}},
\\[0.5em]
\mathrm{PQ}_{4}^{F_1}
=
F_{1,\mathrm{micro}}.
\end{gathered}
\]

Micro averaging gives greater weight to competency questions with larger answer sets.
Failed queries contribute no returned answers, while their gold-standard answers remain included when calculating completeness.

\paragraph{PQ5: Exact CQ Satisfaction}
We additionally report exact CQ satisfaction, \(\mathrm{PQ}_{5}=\operatorname{CQExact}\), as the proportion of competency questions for which the returned answer set exactly matches the gold-standard answer set.
Failed queries count as non-matches.

\paragraph{PQ6: Hierarchy-Aware CQ Answer Quality}
For CQs whose answers are ontology terms, binary answer matching is replaced by the hierarchy-based \(\operatorname{hscore}(a,g)\) defined above.
Predicted and gold answers are aligned through maximum-weight one-to-one matching, from which hierarchy-aware correctness,
completeness, and \(F_1\) are calculated.
We report their macro-averaged \(F_1\) as \(\mathrm{PQ}_{6}=HF_{1,\mathrm{macro}}\).

\paragraph{Stratified Evaluation}
To account for differences in CQ difficulty, we additionally report macro-\(F_1\) stratified by gold-answer cardinality, query complexity, mapping challenge, and reasoning requirement, following analyses such as \cite{orogat2021cbench}.
For each category \(c\), the score is computed over its associated competency questions \(Q_c\):
\[
F_{1,\mathrm{macro}}^{(c)}
=
\frac{1}{|Q_c|}
\sum_{q \in Q_c} F_1(q).
\]
Where mutually exclusive difficulty levels are defined, their equally weighted mean is reported as a secondary balanced score.

\subsection{Implementation}
---

\section{Datasets and Curation}
\label{sec:data}
This benchmark contains 10 hand-curated, real world datasets from the domains of ecology, biodiversity research, genomics \& metabolomics, copolymer chemistry, and history.
In this section, we present an overview for each dataset, and refer to the appendix/our GitHub repository for complete documentation, including complete KG schemas, reference mappings, CQs and gold answers, and target ontology information.

\section{Evaluation}
\label{sec:eval}
---

\section{Discussion}
\subsection{Implications and Limitations}
Our proposed evaluation pipeline design combined with the input files available in the provided benchmark datasets allows not only the concurrent evaluation of KG mappings and the pragmatic usefulness these mappings imply, but also allows the evaluation of varying input scenarios.
In addition to the required inputs (data, CQs, CQ answers, and target ontologies), systems are able to quantify the impact that optional context files (data intent sheets, papers/texts for context, ontology tutorials) and combinations thereof have on KG creation tasks, resulting in new insights regarding what context is especially important for candidate systems.
This may be especially revealing for comparisons of different sets of input CQs: How does CQ quality impact KG mapping/construction task performance?

Additionally, the modular framework design and the generalized representations employed by BLINKG's tabular mapping predictions and our CQ Query Generator module enable the evaluation of systems with varying strategies.
This is emphasized further as systems can skip the mapping task entirely and start evaluation with a materialized KG (Fig. \ref{fig:1}, Stage 3), for example to conduct a CQ evaluation, test consistency and completeness, or validate the resulting KG schema by leveraging our provided schema-extractor module.
This contributes a significant comparison axis for a yet unsolved KG evaluation problem: Two KGs may use different schemas and ontologies, but solve downstream tasks equally well.
Through still providing gold standard mappings, we are not able to solve the general semantic evaluation of mapping choices for an unrestricted set of available ontologies, but our pipeline enables pragmatic comparison of different graphs, and our benchmark datasets may in future be extended to contain multiple reference mappings.

Regarding error sources, our evaluation framework design has two main implications:
(1) First, we use competency questions as the main feature for pragmatic evaluation.
Therefore, for both our provided benchmark datasets, but also for deployment scenarios, the quality of provided CQs regarding for example their coverage of the input data have significant influence on evaluation scores and generated KGs.
However, this also presents an opportunity as research regarding CQ authoring, generation, and evaluation are gaining traction.
(2) Second, we isolate the mapping task from mapping authoring itself. Through the pipeline's compiler that translates BLINKG's tabular mappings to RML mappings, we eliminate syntactically incorrect RML mappings as an error source.

Regarding limitations, as detailed in the BLINKG paper \cite{castedo2026blinkg}, we are not able to address all open issues:
Memorization and data leakage present both upsides and downsides.
Ontologies and their documentation may leak into LLM's training data, presenting a confounder that influences mapping task performance.
While one could argue that this represents a realistic deployment scenario in which ontologies are in parts known to automated systems, the issue remains that lesser-known ontologies may be left underrepresented.
Additionally, our benchmark datasets may not exhaustively represent real-world domains and scenarios, especially as it is difficult to gauge the difficulty of input data itself, the amount of target ontologies per dataset, and the complexity of the reference mappings itself.
Finally, as our benchmark aims to reflect real-world conditions, the amount of provided context may lead to enhanced task performance, but the computational capabilities and resources needed to leverage this context may be limited between user groups and in open LLMs compared to proprietary models.
\subsection{System Result Discussion}
---

\section{Limitations \& Future Work}
\subsection{Limitations}
---
\subsection{Future Work}
---

\section{Conclusion}
---

\begin{credits}
\subsubsection{\ackname} This research was supported by a Flexpool grant of the
German Centre for Integrative Biodiversity Research Halle-Jena-Leipzig – iDiv.
\subsubsection{Declaration on Generative AI.} During the preparation of this work, the authors used generative AI (ChatGPT) to paraphrase and reword, improve writing style, and for grammar and spelling checks.
 After using the tool, the authors reviewed and edited the content as needed and take full responsibility for the publication’s content.

\subsubsection{\discintname}
The authors have no competing interests to declare that are
relevant to the content of this article.
\end{credits}
%
%
%
\bibliographystyle{splncs04}
\bibliography{lncs.bib}




\end{document}